\documentstyle[11pt,psfig,twoside]{article}
\begin{document}

\newcommand{\dd}{\,{\rm d}}
\newcommand{\ie}{{\it i.e.},\,}
\newcommand{\etal}{{\it et al.\ }}
\newcommand{\eg}{{\it e.g.},\,}
\newcommand{\cf}{{\it cf.\ }}
\newcommand{\vs}{{\it vs.\ }}
\newcommand{\zdot}{\makebox[0pt][l]{.}}
\newcommand{\up}[1]{\ifmmode^{\rm #1}\else$^{\rm #1}$\fi}
\newcommand{\dn}[1]{\ifmmode_{\rm #1}\else$_{\rm #1}$\fi}
\newcommand{\upd}{\up{d}}
\newcommand{\uph}{\up{h}}
\newcommand{\upm}{\up{m}}
\newcommand{\ups}{\up{s}}
\newcommand{\arcd}{\ifmmode^{\circ}\else$^{\circ}$\fi}
\newcommand{\arcm}{\ifmmode{'}\else$'$\fi}
\newcommand{\arcs}{\ifmmode{''}\else$''$\fi}
\newcommand{\MS}{{\rm M}\ifmmode_{\odot}\else$_{\odot}$\fi}
\newcommand{\RS}{{\rm R}\ifmmode_{\odot}\else$_{\odot}$\fi}
\newcommand{\LS}{{\rm L}\ifmmode_{\odot}\else$_{\odot}$\fi}

\newcommand{\Abstract}[2]{{\footnotesize\begin{center}ABSTRACT\end{center}
\vspace{1mm}\par#1\par
\noindent
{~}{\it #2}}}

\newcommand{\TabCap}[2]{\begin{center}\parbox[t]{#1}{\begin{center}
 \small {\spaceskip 2pt plus 1pt minus 1pt T a b l e}
 \refstepcounter{table}\thetable \\[2mm]
 \footnotesize #2 \end{center}}\end{center}}

\newcommand{\TableSep}[2]{\begin{table}[p]\vspace{#1}
\TabCap{#2}\end{table}}

\newcommand{\FigCap}[1]{\footnotesize\par\noindent Fig.\ %
 \refstepcounter{figure}\thefigure. #1\par}

\newcommand{\TableFont}{\footnotesize}
\newcommand{\TableFontIt}{\ttit}
\newcommand{\SetTableFont}[1]{\renewcommand{\TableFont}{#1}}

\newcommand{\MakeTable}[4]{\begin{table}[htb]\TabCap{#2}{#3}
 \begin{center} \TableFont \begin{tabular}{#1} #4 
 \end{tabular}\end{center}\end{table}}

\newcommand{\MakeTableSep}[4]{\begin{table}[p]\TabCap{#2}{#3}
 \begin{center} \TableFont \begin{tabular}{#1} #4 
 \end{tabular}\end{center}\end{table}}

\newenvironment{references}%
{
\footnotesize \frenchspacing
\renewcommand{\thesection}{}
\renewcommand{\in}{{\rm in }}
\renewcommand{\AA}{Astron.\ Astrophys.}
\newcommand{\AAS}{Astron.~Astrophys.~Suppl.~Ser.}
\newcommand{\ApJ}{Astrophys.\ J.}
\newcommand{\ApJS}{Astrophys.\ J.~Suppl.~Ser.}
\newcommand{\ApJL}{Astrophys.\ J.~Letters}
\newcommand{\AJ}{Astron.\ J.}
\newcommand{\IBVS}{IBVS}
\newcommand{\PASP}{P.A.S.P.}
\newcommand{\Acta}{Acta Astron.}
\newcommand{\MNRAS}{MNRAS}
\renewcommand{\and}{{\rm and }}
\section{{\rm REFERENCES}}
\sloppy \hyphenpenalty10000
\begin{list}{}{\leftmargin1cm\listparindent-1cm
\itemindent\listparindent\parsep0pt\itemsep0pt}}%
{\end{list}\vspace{2mm}}

\def\TYLDA{~}
\newlength{\DW}
\settowidth{\DW}{0}
\newcommand{\dw}{\hspace{\DW}}

\newcommand{\refitem}[5]{\item[]{#1} #2%
\def\REFARG{#3}\ifx\REFARG\TYLDA\else, {\it#3}\fi
\def\REFARG{#4}\ifx\REFARG\TYLDA\else, {\bf#4}\fi
\def\REFARG{#5}\ifx\REFARG\TYLDA\else, {#5}\fi.}

\newcommand{\Section}[1]{\section{#1}}
\newcommand{\Subsection}[1]{\subsection{#1}}
\newcommand{\Acknow}[1]{\par\vspace{5mm}{\bf Acknowledgements.} #1}
\pagestyle{myheadings}

\def\thefootnote{\fnsymbol{footnote}}
\begin{center}
{\Large\bf The Optical Gravitational Lensing Experiment.\\
\vskip3pt
Cepheids in the Magellanic Clouds.\\
\vskip3pt
I. Double-Mode Cepheids\\
\vskip3pt
in the Small Magellanic Cloud\footnote{Based on observations obtained with the 
1.3~m Warsaw telescope at the Las Campanas Observatory of the Carnegie 
Institution of Washington.}} 
\vskip1cm
{\bf
A.~~U~d~a~l~s~k~i$^1$,~~I.~~S~o~s~z~y~{\'n}~s~k~i$^1$,
~~M.~~S~z~y~m~a~{\'n}~s~k~i$^1$,~~M.~~K~u~b~i~a~k$^1$,
~~G.~~P~i~e~t~r~z~y~\'n~s~k~i$^1$,
~~P.~~W~o~\'z~n~i~a~k$^2$,~~ and~~K.~~\.Z~e~b~r~u~\'n$^1$}
\vskip3mm
{$^1$Warsaw University Observatory, Al.~Ujazdowskie~4, 00-478~Warszawa, Poland\\
e-mail: (udalski,soszynsk,msz,mk,pietrzyn,zebrun)@sirius.astrouw.edu.pl\\
$^2$ Princeton University Observatory, Princeton, NJ 08544-1001, USA\\
e-mail: wozniak@astro.princeton.edu}
\vskip5mm
\end{center}

\Abstract{We present a sample of 93 double-mode Cepheids detected in the
2.4  square degree area in the central part of the SMC. 23 stars from
the sample  pulsate in the fundamental mode and the first overtone while
70 objects are  the first and second overtone pulsators. This is the
largest sample of such  type Cepheids detected in one environment so
far. 

We analyze period ratio of double-mode Cepheids and Fourier parameters
of  decomposition of the light curves of these objects. We also present
location of different type Cepheids from the SMC in the color-magnitude
diagram and show their distribution of ${V-I}$ color indices. We find
one object which  is probably a blend, either physical or optical, of
two Cepheids pulsating in  the fundamental mode.}{~}

\Section{Introduction} 
Microlensing searches for dark matter in the Galaxy provide an unique 
by-product -- huge databases of precise photometric measurements of tens
 million stars in the Magellanic Clouds and dense Galactic fields. The 
measurements span a few years and are ideally suited for variable star
study.  One of the group of variable stars profiting a lot from
microlensing surveys  data are pulsating variable stars. The Magellanic
Clouds offer an ideal sample  of those stars due to their large
population there and approximately the same  distance. Unfortunately,
they have been neglected photometrically for years. 

About 1400 Cepheids were identified in the Large Magellanic Cloud by the
MACHO  microlensing team (Alcock \etal 1995). The majority of objects
are regular  single-mode pulsators. Alcock \etal (1995) also presented a
sample of 45  double-mode Cepheids identified among detected objects.
The double-mode  Cepheids (called sometimes beat Cepheids) pulsate
simultaneously in two radial  modes: either fundamental and first
overtone (FU/FO) or first and second  overtones (FO/SO). The double-mode
Cepheids are relatively rare, only fourteen  were identified in the
Galaxy so far (Pardo and Poretti 1997). Large sample of  double-mode
Cepheids from the LMC, increased later to 75 objects, allowed for  more
detailed study of properties of these objects (Alcock \etal 1999). 

The double-mode Cepheids were also identified in the Small Magellanic
Cloud.  27 object sample was reported by the MACHO team (Alcock \etal
1997) and 11  object sample by the EROS team (Beaulieu \etal 1997). The
sample of the SMC  Cepheids is very important because smaller
metallicity of the SMC allows to  study dependence of pulsation
properties of double-mode Cepheids on their  metallicity. 

The Magellanic Clouds were included to the observing targets of the
Optical  Gravitational Lensing Experiment (OGLE) microlensing search at
the beginning  of the second phase of the project, OGLE-II, in January
1997 (Udalski, Kubiak  and Szyma{\'n}ski 1997). After two years of
constant monitoring of the  Magellanic Clouds, the photometric databases
of the OGLE project are complete  enough to allow for search for
Cepheids in the Magellanic Clouds. In this  paper, first of the series
on Cepheid variable stars in the Magellanic Clouds,  we present results
of the search for double-mode Cepheids in the central  regions of the
SMC leading to discovery of 93 such objects -- the largest  sample of
double-mode Cepheids detected so far. 

\Section{Observations}
All observations presented in this paper were carried out during the
second  phase of the OGLE experiment with the 1.3-m Warsaw telescope at
the Las  Campanas Observatory, Chile, which is operated by the Carnegie
Institution of  Washington. The telescope was equipped with the "first
generation" camera with  a SITe ${2048\times2048}$ CCD detector working
in the drift-scan mode. The  pixel size was 24~$\mu$m giving the 0.417
arcsec/pixel scale. Observations of  the SMC were performed in the
"slow" reading mode of the CCD detector with the  gain 3.8~e$^-$/ADU and
readout noise about 5.4~e$^-$. Details of the  instrumentation setup can
be found in Udalski, Kubiak and Szyma{\'n}ski  (1997). 

Observations of the SMC started on June~26, 1997. As the microlensing
search  is planned to last for a few years, observations of selected
fields will be  continued during the following seasons. In this paper we
present data  collected up to March~4, 1998. Observations were obtained
in the standard {\it  BVI}-bands with majority of measurements made in
the {\it I}-band. 

Photometric data collected during the first observing season of the SMC
for  11 fields (SMC$\_$SC1--SMC$\_$SC11) covered about 2.4 square degree
of the  central parts of the SMC and were used to construct the {\it
BVI} photometric  maps of the SMC (Udalski \etal 1998a). The reader is
referred to that paper  for more details about methods of data
reduction, tests on quality of  photometric data, astrometry, location
of observed fields etc. 

\Section{Selection of Double-Mode Cepheids}
The search for variable objects in 11 SMC fields was performed using 
observations in the {\it I}-band in which majority of observations was 
obtained. Typically about 160--200 epochs were available for each
analyzed  object with the lower limit set to~50. The mean {\it I}-band
magnitude of  objects was limited to ${I<20}$~mag. Candidates for
variable stars were  selected based on comparison of the standard
deviation of all individual  measurements of a star with typical
standard deviation for stars of similar  brightness. Light curves of
selected candidates were then searched for  periodicity using the AoV
algorithm (Schwarzenberg-Czerny 1989). 

Candidates for Cepheids were selected from the entire sample of variable
stars  based on visual inspection of the light curves and location in
the  color-magnitude diagram (CMD) within the area limited by
${I{<}18.5}$~mag and  ${0.25{<}(V{-}I){<}1.3}$~mag. Several objects
located outside this region but with  evident Cepheid light curves were
also included to this sample (\eg highly  reddened Cepheids). In total
more than 2300 Cepheid candidates were found in  the 2.4 square degree
area of the SMC bar. The catalog of all objects will be  presented in
the following papers of this series. 

Selection of double-mode Cepheids was performed in two stages. First, in
the  preliminary search, we used results of the general variable star
search  described above. The mean light curve folded with the AoV period
of each  Cepheid candidate was fitted by high order polynomial and
subtracted from the  light curve. Double-mode Cepheids usually fold with
the higher amplitude  periodicity displaying abnormally large scatter in
the light curve. The  residuals were then searched for periodic signal
and, if detected, such a  candidate was marked for further analysis.
Then a histogram of the ratio of  the shorter to the longer period of
selected double-mode Cepheid candidates  was constructed. It exhibited
two clear sharp peaks corresponding to the ratio of  the first overtone
to the fundamental period, ${\approx0.735}$, and the second  to the
first overtone period, ${\approx0.805}$, in good agreement with Alcock 
\etal (1997). The list of the double-mode Cepheid candidates from this 
search included stars having the period ratio within ${\pm0.02}$ from
these  values -- the range wide enough to avoid missing potential
outliers.

The second, final search for double-mode Cepheids was performed using
the {\sc  Clean} algorithm of period determination (Roberts, Leh{\'a}r
and Dreher 1987).  All 2300 objects from the Cepheid candidate list were
subjected to the {\sc  Clean} period analysis. Having well established
limits for the period ratio of  double-mode Cepheids from the
preliminary analysis, only those objects which  exhibited suitable
period ratio (${\pm0.015}$) between the highest peak in the  power
spectrum and one of the next four strongest peaks were further analyzed.
The final list of the double-mode Cepheid candidates presented in this
paper was obtained after  careful visual inspection of the {\sc Clean}
power spectra of each object.

\Section{Double-Mode Cepheids in the SMC}

Double-mode Cepheids detected in the central area of the SMC are listed
in  Tables~1 and 2. 95 objects were detected but 93 of them are unique.
Two stars  are located in the overlapping regions between fields and
they were discovered  independently in each field. Table~1 contains
systems which pulsate in the  fundamental and first overtone modes while
Table~2 -- objects pulsating in the  first and second overtones. Basic
parameters of each star: right ascension and  declination (J2000), the 
intensity-mean {\it I}-band magnitude, ${(B-V)}$ and ${(V-I)}$ colors,
both  periods and their ratio are provided. Accuracy of periods is 
about  ${7\cdot10^{-5}P}$. Finding charts for all objects are presented
in   Appendix~A. The size of the {\it I}-band subframes is
${60\times60}$ arcsec;   North is up and East to the left. 
\begin{figure}[p]
\psfig{figure=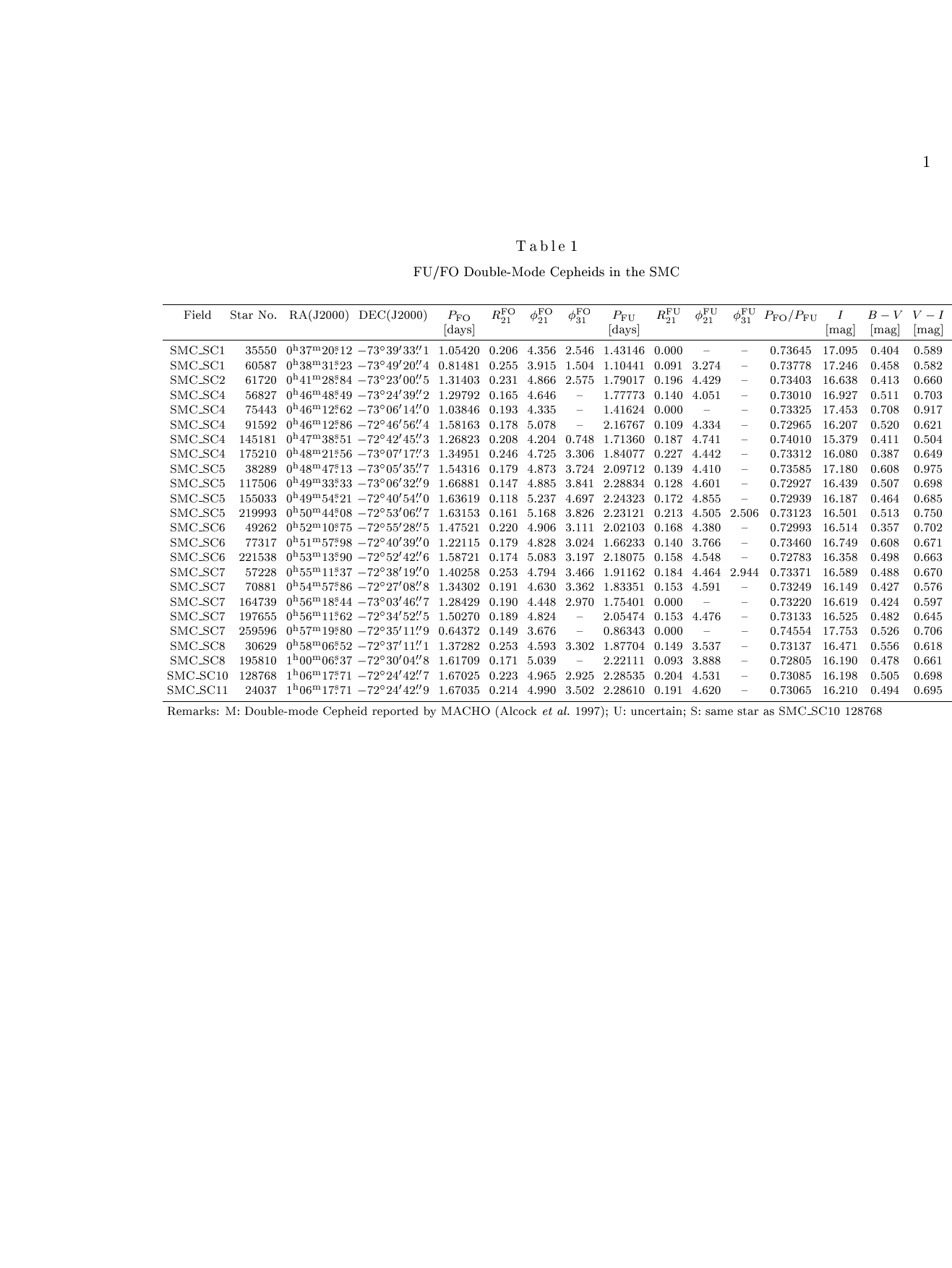,bbllx=100pt,bblly=325pt,bburx=670pt,bbury=690pt,angle=90,clip=}
\end{figure}
\begin{figure}[p]
\psfig{figure=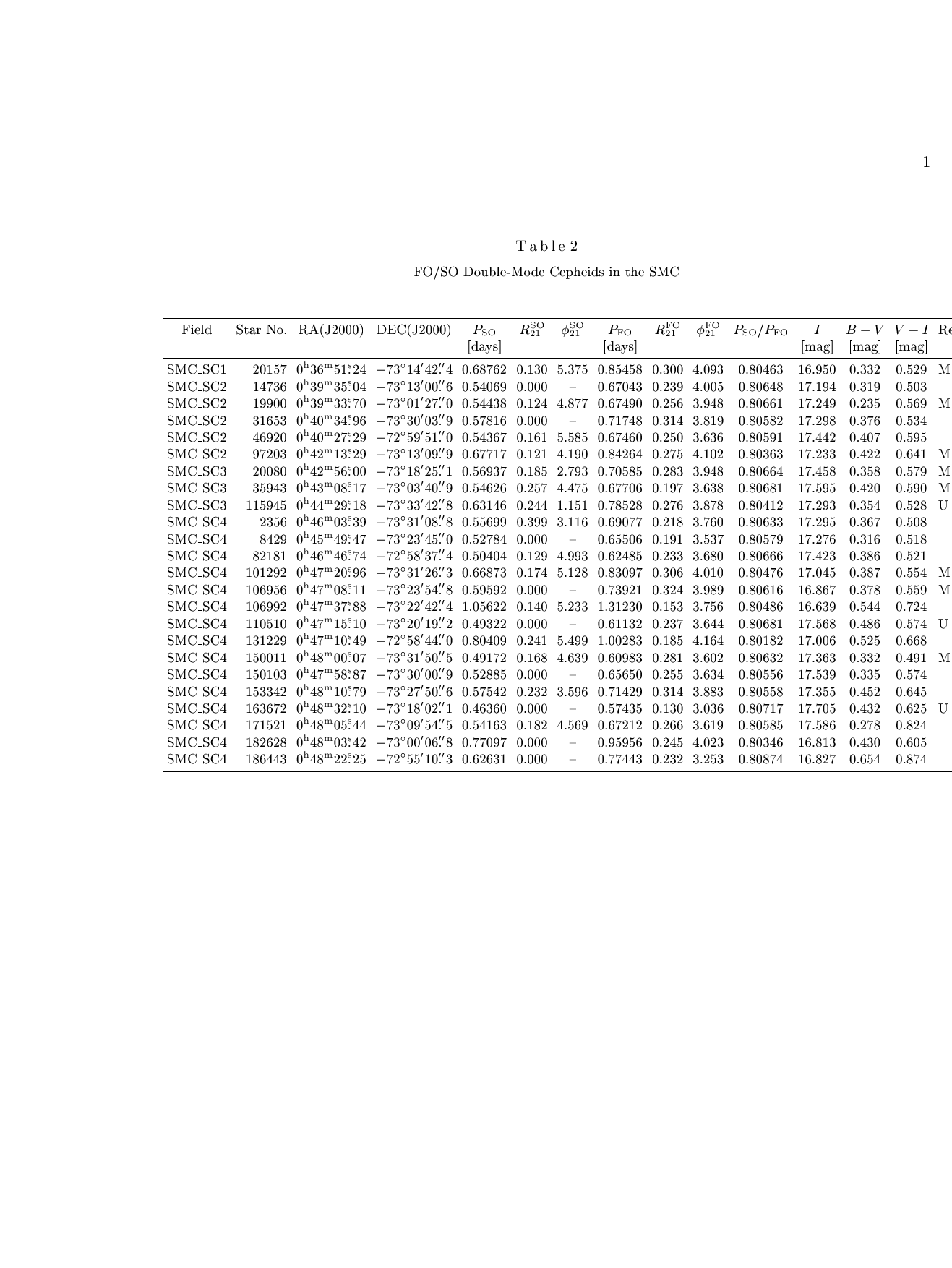,bbllx=100pt,bblly=325pt,bburx=670pt,bbury=690pt,angle=90,clip=}
\end{figure}
\begin{figure}[p]
\psfig{figure=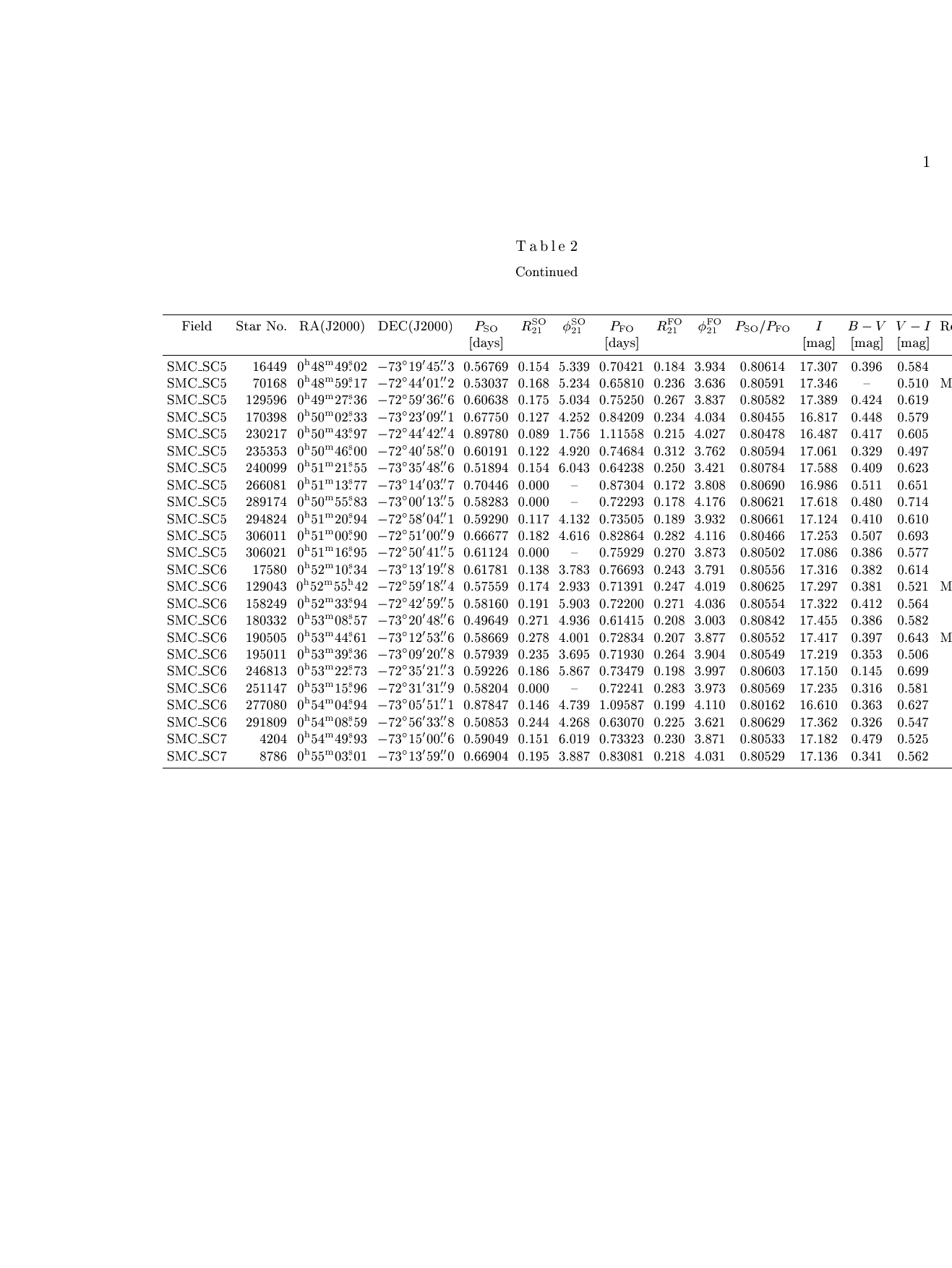,bbllx=100pt,bblly=325pt,bburx=670pt,bbury=690pt,angle=90,clip=}
\end{figure}
\begin{figure}[p]
\psfig{figure=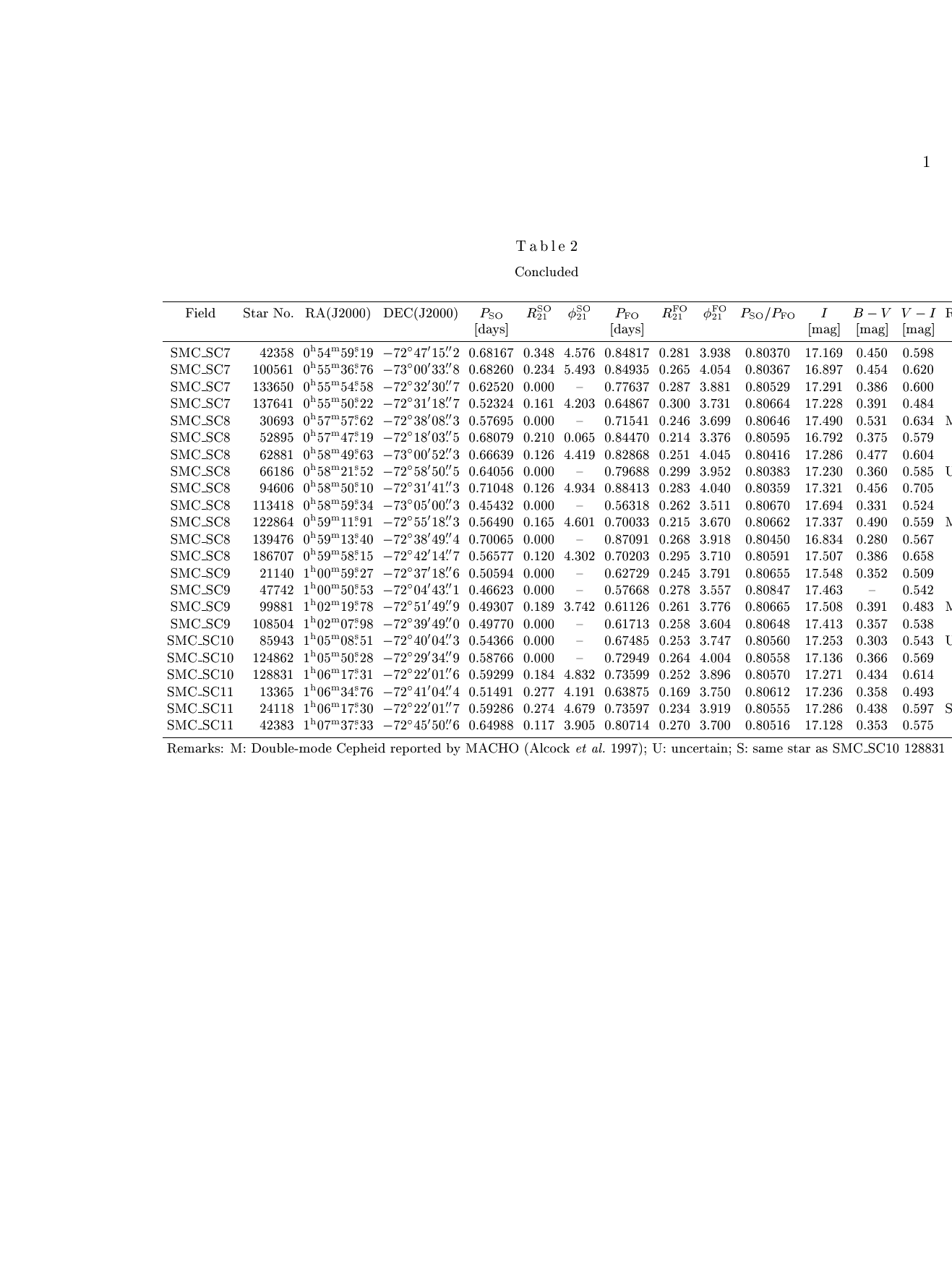,bbllx=100pt,bblly=325pt,bburx=670pt,bbury=690pt,angle=90,clip=}
\end{figure}

Appendices~B and C show the light curves of FU/FO and FO/SO pulsators, 
respectively. The first and second columns in each Appendix contain
original  photometric data folded with the shorter and longer periods
while the  remaining columns show variability attributed to each mode
after subtraction  of the other period variability approximated by
Fourier series of fifth order.  For objects revealing also periodicity
equal to the sum and/or difference of both  mode frequencies and having
an amplitude larger than twice the formal error --  such terms were also
subtracted from the original data. {\it BVI} photometry  of all objects
will be available from the OGLE Internet archive when the  catalog of
Cepheids in the SMC is released. 

Completeness of the sample is determined by completeness of the variable
star  search in the OGLE databases and efficiency of double-mode Cepheid
detection  algorithm. Completeness of the OGLE variable stars catalog
was already  estimated for eclipsing stars which are much more difficult
to detect. For  objects brighter than ${I=17}$ it is likely to be higher
than 90\% (Udalski  \etal 1998b). For stars as bright and easy to detect
as Cepheids it should be  similar or even higher. Completeness of the
detection algorithm can be  assessed by comparison of results obtained
in the preliminary and final ({\sc  Clean}) searches. More than 90\%
objects in both lists are common suggesting  good completeness of the
search. 

As a test of completeness we cross-identified double mode Cepheids
reported by  MACHO (Alcock \etal 1997). 19 out of 20 objects which are
located in the OGLE  fields were detected during our search. They are
marked by letter 'M' in the  last column of Tables~1 and 2. The
remaining object,  MACHO*00:57:27.0-73:04:39, was also analyzed but it
seems to be a single  period Cepheid. 

\Section{Discussion}
93 double-mode Cepheids were identified during the presented search in
the 2.4  square degree area in the central bar of the SMC. 23 objects
pulsate  simultaneously in the fundamental mode and first overtone while
70 objects in  the first and second overtones. The sample constitutes
the most numerous  sample of double-mode Cepheids located in one
environment. Completeness of the  sample is high, likely larger than
90\%. 

\Subsection{Period Ratio in Double-Mode Cepheids}

Fig.~1 presents the ratio of periods of the FU/FO and FO/SO pulsators
plotted  as a function of the lower mode period. In both cases a clear
dependence on  the period is seen, similarly to the Galactic and LMC
Cepheids (Alcock \etal  1995). More numerous sample allows for more
precise approximation of that  dependence. 
\begin{figure}[htb]
\vskip-5mm
\centerline{\psfig{figure=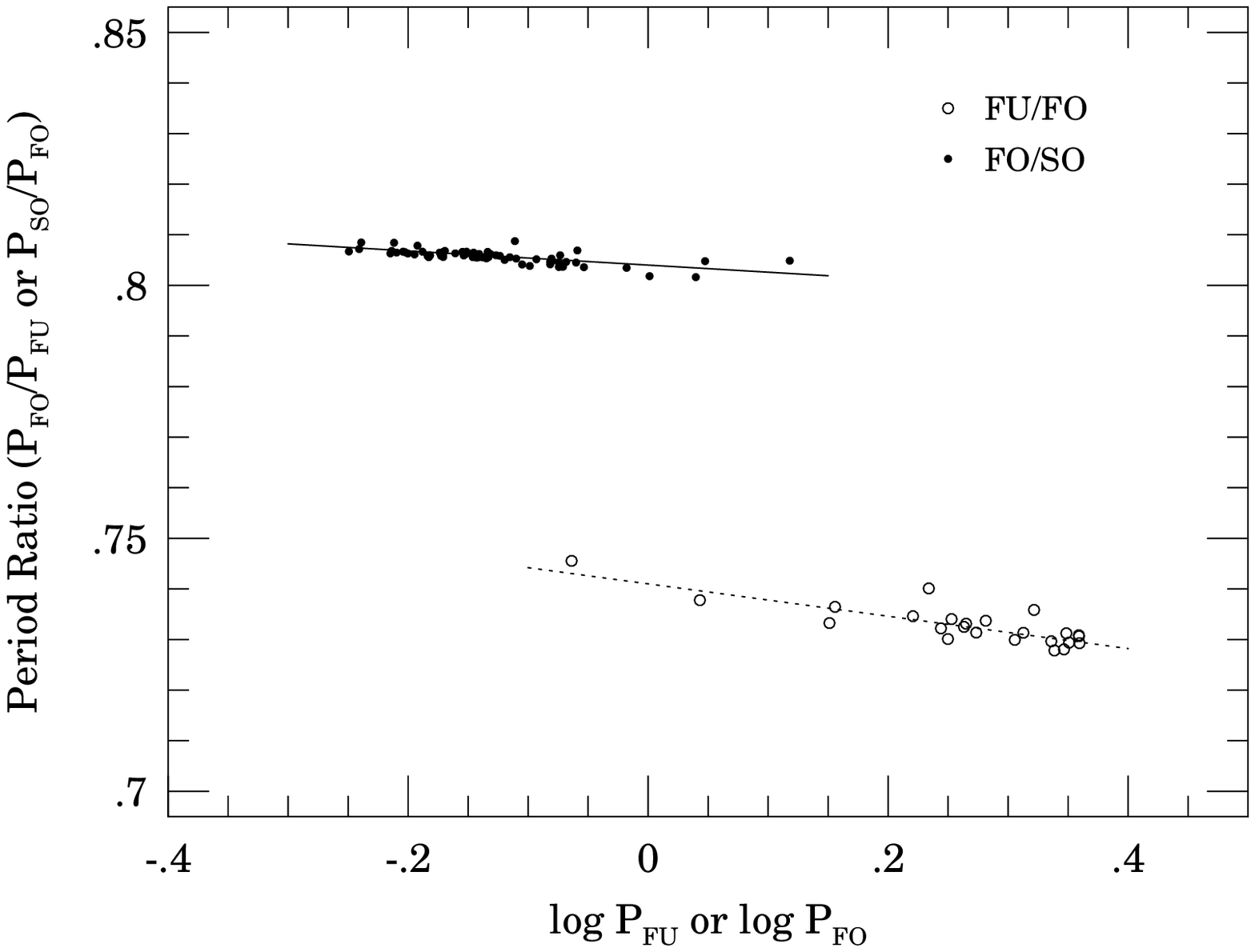,bbllx=25pt,bblly=40pt,bburx=510pt,bbury=410pt,width=12cm,clip=}}
\FigCap{Ratio of periods in double-mode Cepheids plotted as a function
of the longer period. Dotted and solid lines mark the best linear fits
given by Eqs.~1 and 2, respectively.} 
\end{figure}

\noindent
The best linear fits are as follows: 

\noindent
FU/FO Cepheids:

\begin{tabular}{r@{\hspace{5pt}}c@{\hspace{5pt}}r@{\hspace{5pt}}c@{\hspace{5pt}}l}
$\phantom{XXXXXXXX}
P_{\rm FO}/P_{\rm FU}$ & $=$ & 0.741 & $-$ & $0.032\times \log P_{\rm FU}$\,,\\
\noalign{\vskip-5pt}
\noalign{\rightline{(1)~~~~~~~}}
\noalign{\vskip-5pt}
                        &     & 0.001 &     &  0.005\\ 
\end{tabular}

\noindent
FO/SO Cepheids:

\begin{tabular}{r@{\hspace{5pt}}c@{\hspace{5pt}}r@{\hspace{5pt}}c@{\hspace{5pt}}l}
$\phantom{XXXXXXXX}
P_{\rm SO}/P_{\rm FO}$ & $=$ & 0.804 & $-$ & $0.014\times \log P_{\rm FO}$\,.\\
\noalign{\vskip-5pt}
\noalign{\rightline{(2)~~~~~~~}}
\noalign{\vskip-5pt}
                        &     & 0.001 &     &  0.002\\ 
\end{tabular}

Comparison of coefficients in Eqs.~(1) and (2) with similar ones for the 
Galactic and LMC double-mode Cepheids:

$$P_{\rm FO}/P_{\rm FU}=0.733-0.034\times \log P_{\rm FU}$$
$$P_{\rm SO}/P_{\rm FO}=0.803-0.022\times \log P_{\rm FO}$$

\noindent
for the LMC Cepheids and

$$P_{\rm FO}/P_{\rm FU}=0.720-0.027\times \log P_{\rm FU}$$

\noindent
for the Galactic Cepheids (Alcock \etal 1995) indicates that the ratio
of  periods for  FU/FO pulsators is slightly larger for the SMC Cepheids
than for  the LMC and  Galactic objects which are more metal rich: ${\rm
[Fe/H]}=-0.7$  for the SMC Ce\-pheids \vs ${\rm [Fe/H]}=-0.3$ and ${\rm
[Fe/H]}=0.0$ for the  LMC and Galactic Ce\-pheids, respectively. On the
other hand the ratio for   FO/SO Ce\-pheids is almost identical with the
ratio of the LMC objects but the   slope of the relation is flatter. 

\Subsection{Fourier Decomposition of Light Curves of Double-Mode Cepheids}
Fourier decomposition of light curves of pulsating stars has been widely
used  for analyzing their properties (Simon and Lee 1981). In the case
of Cepheids  the ratio of amplitudes of the first harmonic and the
fundamental period,  ${R_{21}=A_2/A_1}$, and phase difference,
${\phi_{21}=\phi_2-2\phi_1}$ are  particularly useful. Both allow to
distinguish between the fundamental mode  and first overtone pulsators.
The $R_{21}$ \vs $\log P$ diagram constructed  for about 1400 Cepheids
from the LMC (Alcock \etal 1999) shows two distinct  and well separated
"V-shape" sequences for Cepheids pulsating in the  fundamental mode and
the first overtone. In the similar diagram $\phi_{21}$  \vs $\log P$ the
sequences for both modes of pulsation are also well defined  but the
separation is smaller and in some ranges of periods they overlap. 

Fig.~2 presents the $R_{21}$ \vs $\log P$ and $\phi_{21}$ \vs $\log P$ 
diagrams constructed for about 2300 single-mode Cepheids (small dots)
from the  SMC. The Ce\-pheids come from the preliminary catalog of
Cepheids in the SMC,  and therefore some very tiny contamination by
non-Cepheid variable stars is  still possible in these diagrams.
Nevertheless both diagrams look basically  the same as for the LMC
Cepheids with well-separated "V-shape" sequences in  the $R_{21}$ \vs
$\log P$ diagram and two characteristic sequences in the  $\phi_{21}$
\vs $\log P$ diagram. 

\begin{figure}[p]
\psfig{figure=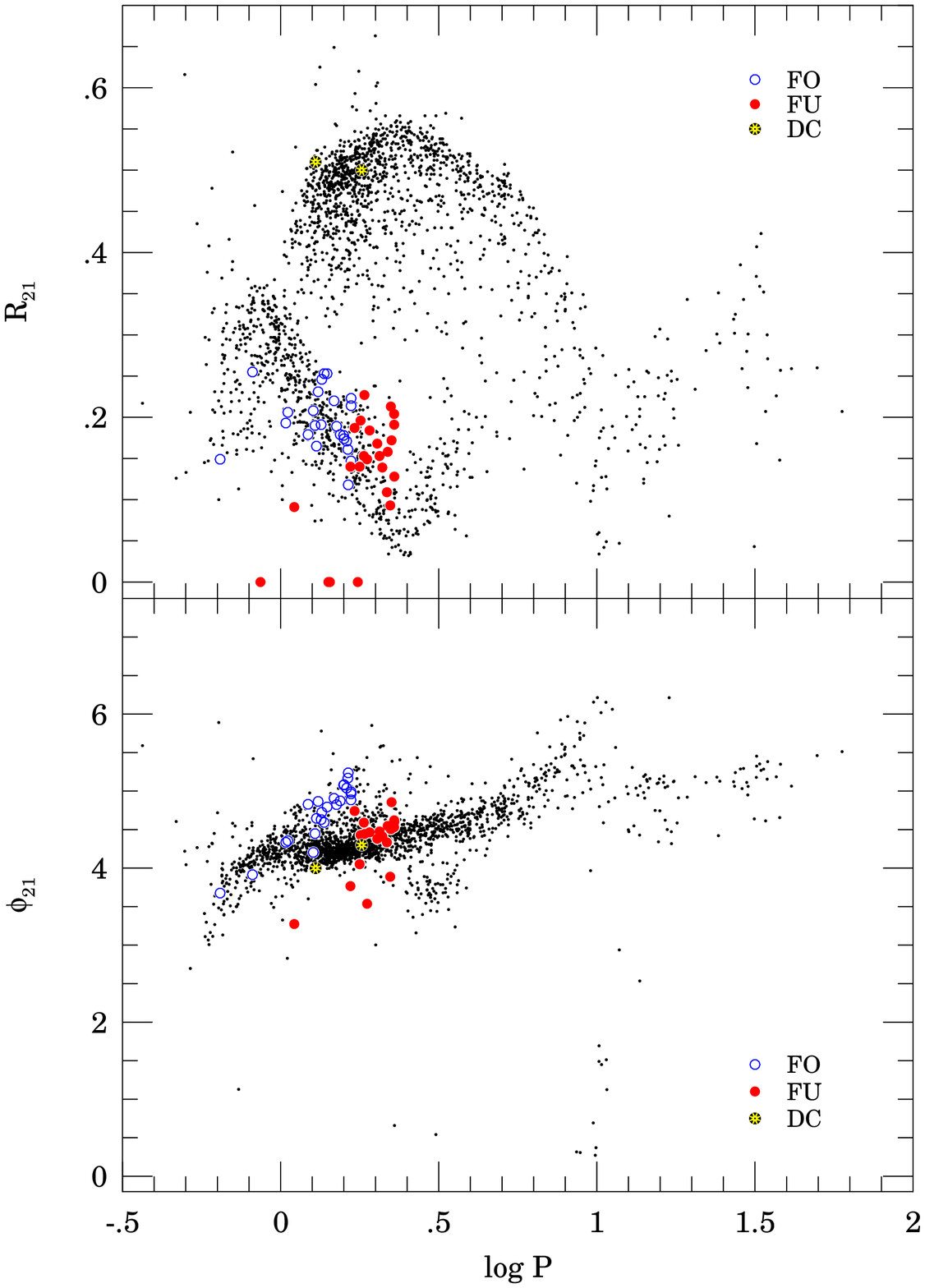,bbllx=30pt,bblly=40pt,bburx=505pt,bbury=705pt,width=12.5cm,clip=}
\FigCap{${R_{21}}$ and ${\phi_{21}}$ \vs $\log P$ diagrams for single-mode
Cepheids from the SMC (small dots). Large open and filled circles mark
values of the first overtone and fundamental mode pulsations in the FU/FO
double-mode Cepheids, respectively. Star symbols denote values of
${R_{21}}$ and ${\phi_{21}}$ for "double Cepheid", SMC$\_$SC5 208044.}
\end{figure}
\begin{figure}[htb]
\vspace*{-9pt}
\psfig{figure=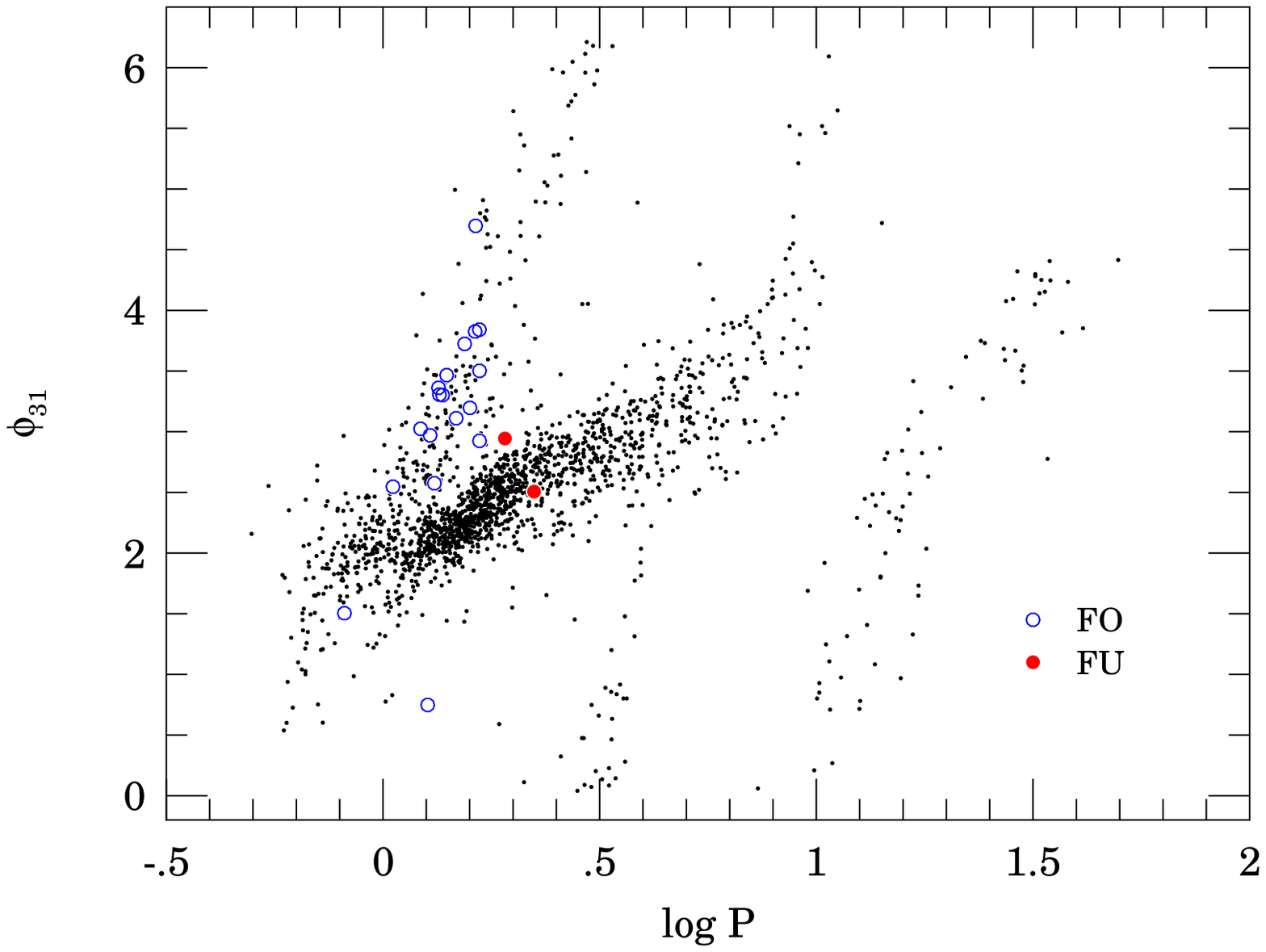,bbllx=30pt,bblly=40pt,bburx=505pt,bbury=405pt,width=12.5cm,clip=}
\vspace*{-3pt}
\FigCap{${\phi_{31}}$ \vs ${\log P}$ diagram for single-mode
Cepheids from the SMC (small dots). Large open and filled circles mark 
values of the first overtone and fundamental mode pulsations in the FU/FO
double-mode Cepheids, respectively.} 
\end{figure}
To check behavior of double-mode Cepheids we decomposed their light
curves to  the sum of two Fourier series of fifth order corresponding to
both  periodicities including the terms of sum and difference of mode
frequencies  when their amplitudes were larger than twice the formal
errors. Then we  calculated $R_{21}$ and $\phi_{21}$ for both pulsating
modes. They are listed  in Tables~1 and 2. 

Results for the FU/FO Cepheids are shown in Fig.~2: the values for the
fundamental  mode pulsation are plotted with large filled dots while for
the first overtone  mode with open circles. Objects with non-significant
first harmonic amplitude,  $A_2$, (\ie with almost sinusoidal light
curve) have ${R_{21}=0}$ and their  $\phi_{21}$ is not defined. 

The main conclusion which can be drawn from Fig.~2 is that while the
first  overtone pulsations usually dominate in this class of double-mode
Cepheids and  their $R_{21}$ and $\phi_{21}$ fall in the sequences of
the single-mode first  overtone pulsators, the fundamental mode
pulsations have $R_{21}$ values much  smaller than corresponding
single-mode fundamental mode Cepheids. This means  that the fundamental
mode pulsations in double-mode Cepheids are suppressed making  the light
curve not that sharp and more sinusoidal than for single-mode  Cepheids
of that type. In the most extreme cases the fundamental mode  pulsations
are low amplitude almost sinusoidal variations (see Appendix~B). 

\begin{figure}[p]
\psfig{figure=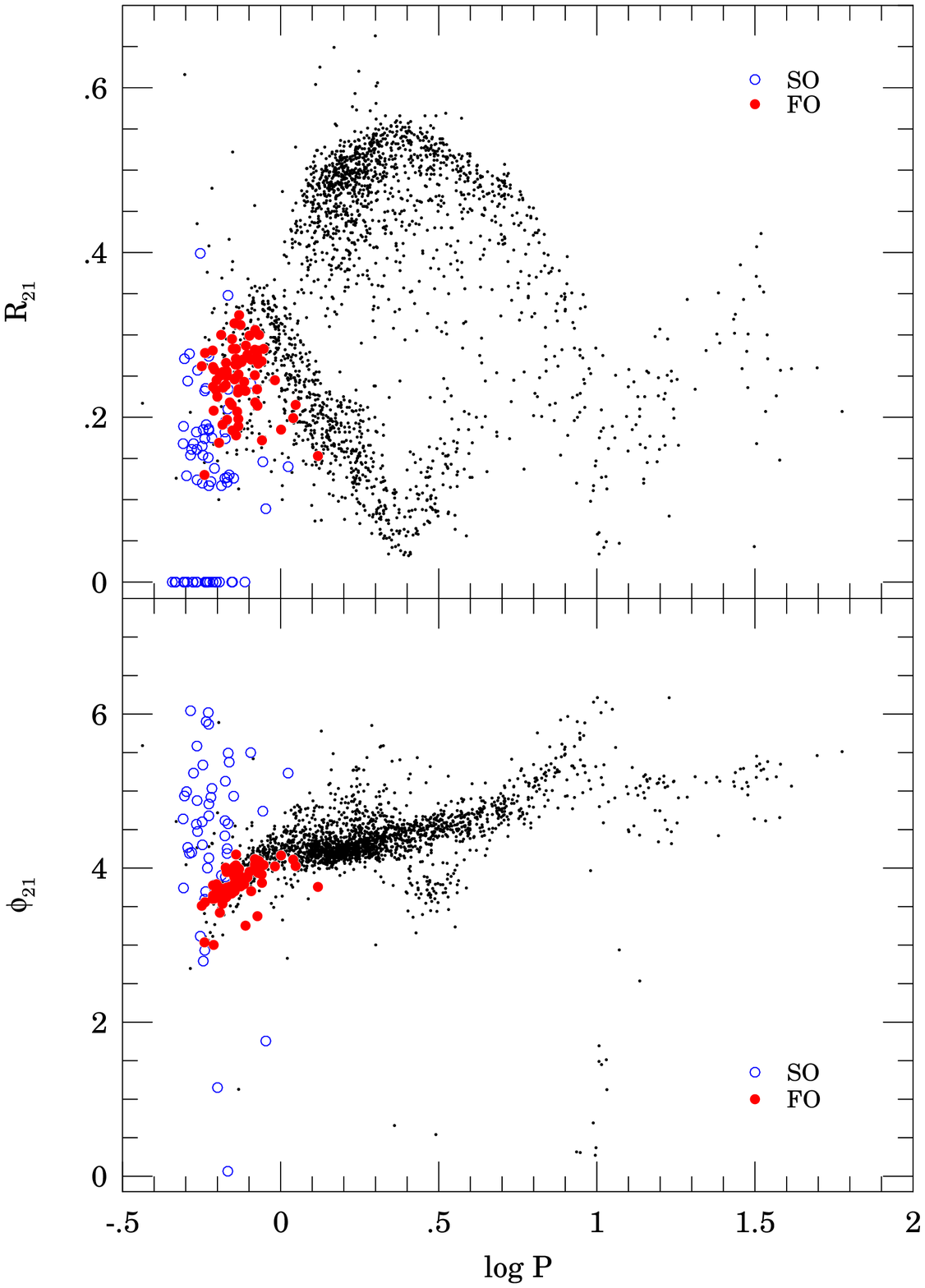,bbllx=30pt,bblly=40pt,bburx=505pt,bbury=705pt,width=12.5cm,clip=}
\FigCap{${R_{21}}$ and ${\phi_{21}}$ \vs ${\log P}$ diagrams for single-mode
Cepheids from the SMC (small dots). Large open and filled circles mark 
values of the second and first overtone pulsations in the FO/SO
double-mode Cepheids, respectively.} 
\end{figure}
$\phi_{21}$ values of the fundamental mode pulsations in double-mode
Cepheids  seem to fall largely in the fundamental mode Cepheid sequence.
Unfortunately,  in this part of the $\phi_{21}$ \vs $\log P$ diagram the
sequences of the fundamental mode and first overtone Cepheids overlap.
Therefore the diagram is not fully conclusive.

Mantegazza and Poretti (1992) suggested that the phase difference of the
second harmonic and fundamental period, ${\phi_{31}=\phi_3-3\phi_1}$
might be better for discrimination between the first overtone and
fundamental mode pulsating objects. The sequences for the first overtone
and fundamental mode Cepheids are better separated in the $\phi_{31}$
\vs $\log P$ than in the $\phi_{21}$ \vs $\log P$ diagram. Therefore we
constructed the $\phi_{31}$ \vs $\log P$ diagram for all single-mode
Cepheids with statistically significant second harmonic amplitude. The
diagram is plotted in Fig.~3. Small dots represent single-mode Cepheids.

Indeed, two well separated sequences starting from approximately  ($\log
P=0, \phi_{31}=2$) are clearly seen. The steeper one is populated by
stars pulsating in the first overtone while the second one, more
horizontal but rising rapidly at (${\log  P=0.9}$, $\phi_{31}=4$) and
better populated, by fundamental mode pulsators. The $\phi_{31}$ values
are periodic with the period equal to $2\pi$ and they are shown in
Fig.~3 in the range $0-2\pi$. Therefore  two additional sequences
starting at (${\log P=0.5}, \phi_{31}=0$) and (${\log P=1.0},
\phi_{31}=0$) are simply continuation of the first overtone and
fundamental mode Cepheid sequences, respectively.

We calculated the $\phi_{31}$ values for all double-mode Cepheids of
FU/FO type with statistically significant second harmonic amplitude. The
values are listed in Table~1. Open circles in Fig.~3 mark positions of
the first overtone pulsation values. As can be seen they fall well in
the single-mode first overtone Cepheid sequence similar to the previous
diagrams. Unfortunately, for the fundamental mode pulsations the
$\phi_{31}$ values could only be derived for two objects from the
sample, SMC$\_$SC5 219993 and SMC$\_$SC7 57228 because the second
harmonic amplitude is non-significant for the remaining Cepheids.
$\phi_{31}$ of these stars are plotted with filled dots in Fig.~3. Both
values are located in the sequence of single-mode Cepheids pulsating in
the fundamental mode. We may, thus,  conclude that the longer period
pulsations in the FU/FO double-mode Cepheids are indeed of the
fundamental mode type in spite of the fact that their $R_{21}$ is
smaller as compared to the values of single-mode objects in the $R_{21}$
\vs ${\log P}$ diagram.

Fig.~4 presents $R_{21}$ \vs $\log P$ and $\phi_{21}$ \vs $\log P$
diagrams  for FO/SO double-mode Cepheids from the SMC. The  first
overtone pulsation values of $R_{21}$ and $\phi_{21}$ are plotted with 
large filled dots while those calculated for the second overtone
pulsations with  open circles. 

The first overtone pulsations also dominate in this group of objects.
Their  $R_{21}$ and $\phi_{21}$ values are located exactly in the
corresponding  sequences of the single-mode first overtone Cepheids. The
second overtone  pulsations are typically low amplitude quasi sinusoidal
light variations (see  Appendix~C). For many objects the first harmonic
amplitude, $A_2$, of the  second overtone period is not statistically
significant (\ie ${R_{21}=0}$),  while for the remaining ones it is
usually very small making $R_{21}$ small,  typically below 0.2. The
$\phi_{21}$ values for the second overtone pulsations  are usually
larger than those corresponding to the first overtone sequence 
providing another way of distinguishing between those two modes of
pulsations. 

\begin{figure}[p]
\psfig{figure=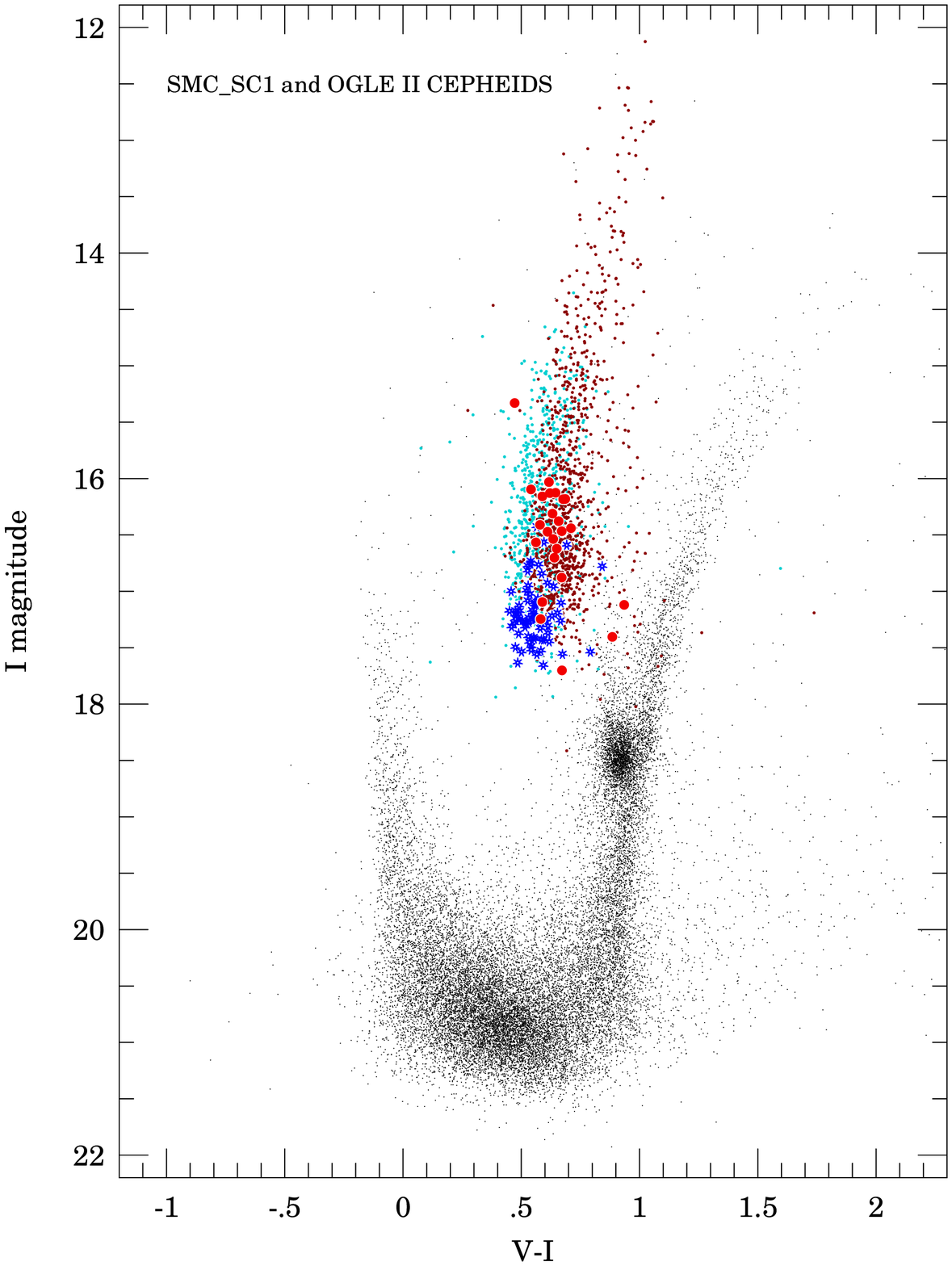,bbllx=35pt,bblly=50pt,bburx=550pt,bbury=730pt,width=13cm,clip=}
\FigCap{Color-magnitude diagram of the SMC$\_$SC1 field. Only about 20\%
of field  stars are plotted by tiny dots. Larger dots show positions of
single-mode  fundamental type Cepheids (darker dots) and first overtone
stars (lighter  dots). Large filled circles and star symbols mark
positions of the FU/FO and FO/SO double-mode Cepheids, respectively.}
\end{figure}
\Subsection{SMC Cepheids in the Color-Magnitude Diagram}
Huge and homogeneous sample of Cepheids from the observed region of the
SMC located practically at the same distance and  relatively small
extinction toward the SMC make it possible to analyze in detail location
of different types of Cepheids in the color-magnitude diagram and  their
distribution of color indices, that  is the temperature distribution of
different mode pulsators. 

We selected four classes of Cepheids: single-mode fundamental and first 
overtone pulsators and double-mode FU/FO and FO/SO objects. The
single-mode  stars were divided into the first overtone and fundamental
mode groups based  on location in the $R_{21}$ \vs $\log P$ and
period-luminosity (${\langle  I\rangle}$ \vs $\log P$) diagrams. These
samples included 705 and 1148  objects, respectively. 

For all objects the intensity-mean magnitudes in the {\it V} and {\it
I}-bands  were calculated. Then the magnitudes of objects from fields  
SMC$\_$SC2--SMC$\_$SC11 were corrected for difference of the mean
reddening   between the field in which a given object is located and the
mean reddening in   the SMC$\_$SC1 field. The mean difference of
reddening between SMC$\_$SC1 and   the remaining fields was derived
based on the mean difference of the  {\it I}-band magnitudes of the red
clump stars in corresponding fields. The   mean {\it I}-band magnitude
of the red clump stars is a good reference of   brightness (Udalski
1998a,b), particularly in homogeneous environment like  central parts of
the  SMC. Similar method was used to determine  map of  extinction of
the Baade's Window in the Galactic bulge by Stanek (1996). 

Thus, magnitudes of all Cepheids were tied to the  reference field
SMC$\_$SC1. The mean extinction of the latter field is 
${A_I=0.11}$~mag, corresponding to $E(B-V)=0.06$ and ${E(V-I)=0.08}$,
based on  the mean {\it I}-band magnitude of the red clump stars,
${I=18.457}$, and the  mean extinction free magnitude of the red clump
stars in the SMC (Udalski  1998a,b). 

Fig.~5 presents the CMD of the field SMC$\_$SC1. Only about 20\% of
field stars were plotted by tiny dots for clarity. Single-mode Cepheids
are plotted with larger dots: fundamental mode -- darker dots, first
overtone -- lighter dots. Positions of FU/FO double-mode Cepheids are
indicated by large filled circles while FO/SO objects by star symbols.

Fig.~5 is a detailed picture of location of the instability strip in the
SMC. The blue edges of the instability strip for both -- fundamental and
first-overtone Cepheids are very well defined. The strip goes almost
vertically up to $I\approx 15$ and bends toward red for brighter objects
-- mostly fundamental mode Cepheids. The red edge is somewhat less sharp
because of several objects reddened more than the mean reddening
correction applied to each field. Nevertheless it can also be precisely
determined.

Double-mode Cepheids are clumped in two distinct locations. The FU/FO
pulsators are on average by about 0.8~mag brighter than FO/SO objects.
They form a vertical sequence in the part of the instability strip
populated by both kinds of single-mode Cepheids: from the blue part of
the instability strip of fundamental mode Cepheids and the red part of
the strip of first overtone objects. The FO/SO pulsators populate
largely low luminosity blue part of the strip of single-mode first
overtone Cepheids. They also form a vertical sequence about 1~mag long
defining the region where such kind pulsations are possible. 

\Subsection{Colors of Double-Mode Cepheids}
More quantitative information on differences of four groups of Cepheids
in the SMC can be obtained by analyzing the distribution of color
indices (\ie temperature distribution).

Fig.~6 shows histograms of $(V-I)$ color indices of all four groups of
Cepheids.  The width of the bin is 0.03~mag. Thick solid and dotted
lines correspond to  the single-mode fundamental and first overtone
Cepheids while thin solid and  dotted lines to the FU/FO and FO/SO
double-mode pulsators. All histograms were  fitted with a Gaussian which
fits well the observed color distributions. In the  case of single-mode 
fundamental mode Cepheids there is a small excess of red  objects caused
by  the redward bending of the instability strip for the  brightest
stars. When we  limit our sample to the objects fainter than $I=15$~mag,
the excess  disappears. 

\begin{figure}[htb]
\psfig{figure=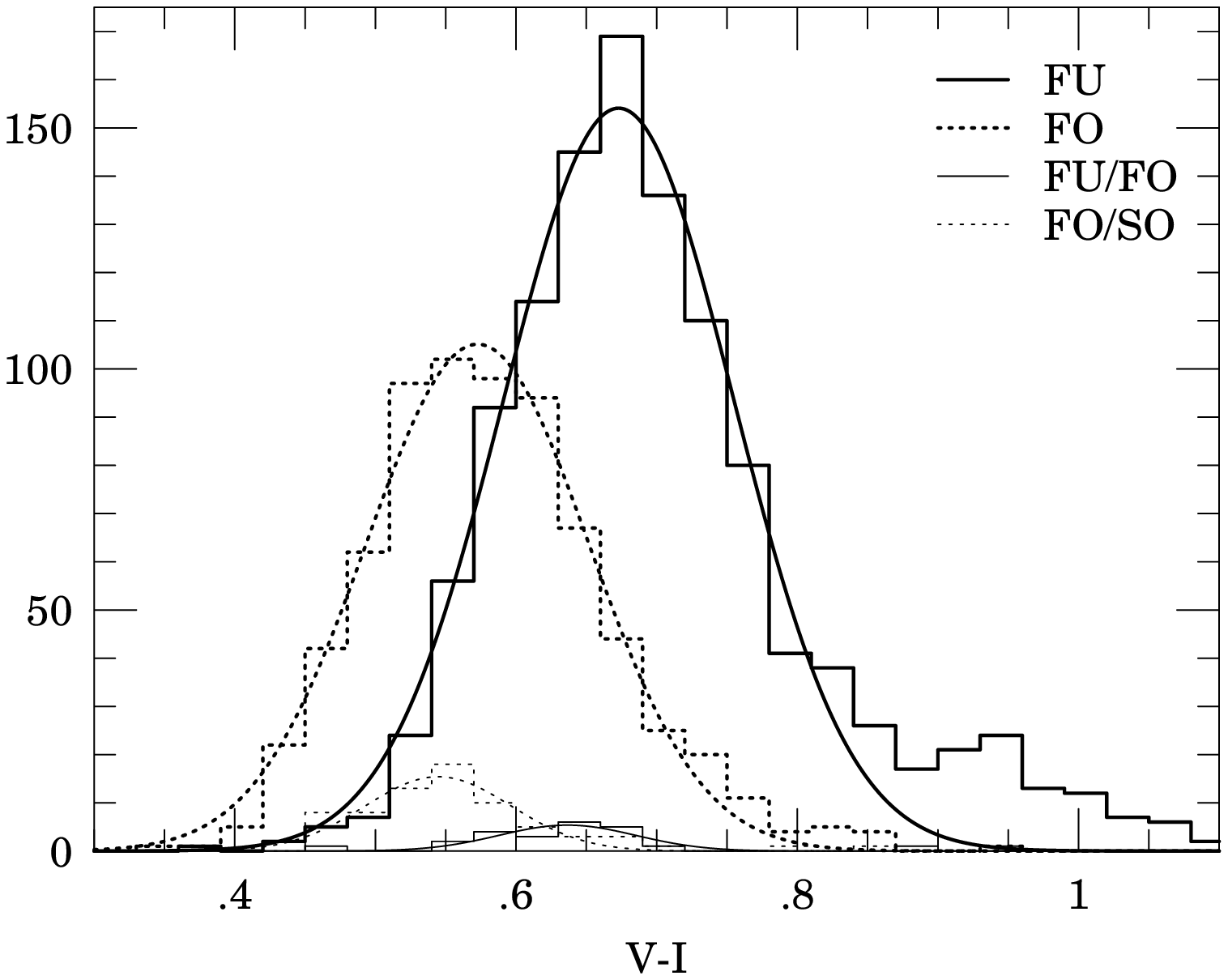,bbllx=50pt,bblly=50pt,bburx=505pt,bbury=410pt,width=12.5cm,clip=}
\FigCap{Histograms of $V-I$ color distribution of single-mode and
double-mode Cepheids in the SMC. Thick lines represent distribution of
single-mode Cepheids: solid line -- fundamental mode pulsators, dotted
line -- first overtone objects. Distribution of double-mode Cepheids is
marked by thin line: solid line -- FU/FO stars, dotted line -- FO/SO
Cepheids. The bins are 0.03~mag wide.}
\end{figure} 
The mean ${V-I}$ color and the standard deviation of its distribution
are 0.673, 0.08 (0.666, 0.08 for $I>15$ sub-sample) and 0.573, 0.08 for
the single-mode fundamental and first overtone  Cepheids, respectively.
For FU/FO and FO/SO double-mode Cepheids the  corresponding values are:
0.634, 0.05 and 0.545, 0.05, respectively. The  single-mode first
overtone Cepheids are on average by about 0.1~mag bluer than 
fundamental mode pulsators. As one could expect the FU/FO double-mode
Cepheids  have ${V-I}$ color distribution in between the first and
fundamental mode  distributions of single-mode stars. The color
distribution of FO/SO double-mode  Cepheids resembles that of the
single-mode first overtone stars but it is  shifted bluewards. 

\Subsection{"Double Cepheid" in the SMC}
The final search with the {\sc Clean} alghoritm was aimed at detection
of  double-mode Cepheids by constraining the searched periods to the
range around  the period ratio of double-mode Cepheids. This approach
omits, however,  potential objects which are a Cepheid blended with
another periodic object.  For instance, three such cases of "double
Cepheids" in the LMC were reported  by Alcock \etal (1995). To look for
similar objects in the SMC we reran our  searching procedure with
another constraint, namely only stars with the second  peak in the power
spectrum (third if the second peak was the harmonic of the  most
prominent one, etc), with frequency outside the double-mode Cepheid
range  and higher than one fourth of the first peak power were selected.

\begin{figure}[htb]
\vspace*{-5mm}
\centerline{\psfig{figure=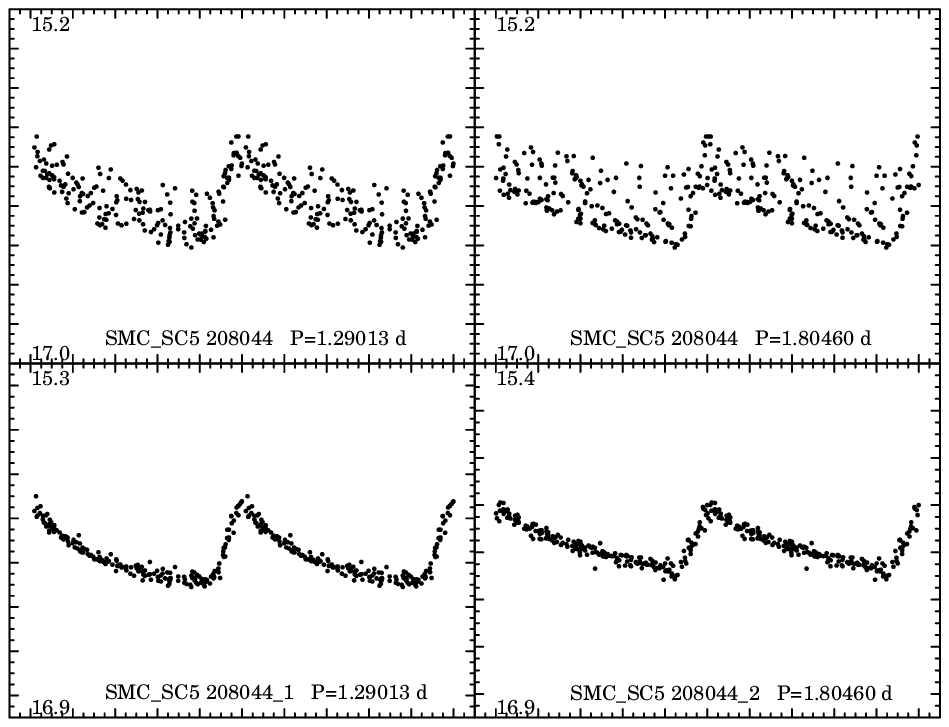,bbllx=120pt,bblly=370pt,bburx=400pt,bbury=600pt,width=9cm,clip=}}
\vspace*{-2mm}
\FigCap{Light curve of "double Cepheid", SMC$\_$SC5 208044. The upper panels
show the original data folded with two detected periods while the lower
panels the light curve of each component after subtraction of
variability of the second star approximated by Fourier series of fifth order.} 
\end{figure}
\begin{figure}[htb]
\centerline{\psfig{figure=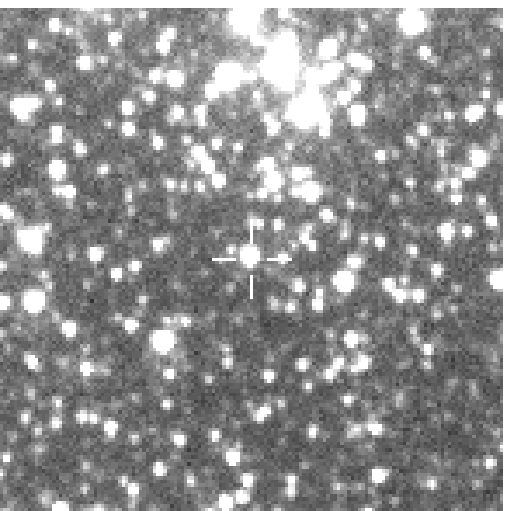,bbllx=0pt,bblly=0pt,bburx=150pt,bbury=150pt,width=4.5cm,clip=}}
\vskip2mm
\FigCap{Finding chart for  "double Cepheid", SMC$\_$SC5 208044.
Size of the {\it I}-band subframe is $60\times 60$ arcsec. North is up
and East to the left.} 
\end{figure}
One object from this sample, SMC$\_$SC5 208044 (${\rm
RA(J2000)}=0\uph50\upm38\zdot\ups78$, ${\rm
DEC(J2000)}=-72\arcd59\arcm02\zdot\arcs1$), seems to consist of two 
blended Cepheids. Fig.~7 shows the light curves of both components of
the  blend. The finding chart is shown in Fig.~8.

Both components seem to be fundamental mode pulsators and although 
their period ratio, 0.715, is close to the FU/FO double-mode Cepheid
range, the shape of the  light curves of both components and Fourier
parameters marked in Fig.~2 by  star symbols leave little doubts that
they are two distinct, blended objects.  Such an interpretation is also
supported by larger brightness of the star than  other objects of
similar period and also lack of periodicity corresponding to  the sum
and difference of two detected frequencies which in a double-mode 
Cepheid of such a large amplitude should be easily detectable. It is 
impossible to conclude whether the system is only an optical blend or 
components are physically bounded. Spectroscopic observations could
provide  additional information to clear this problem. 

\Acknow{We would like to thank Dr.\ W.\ Dziembowski for many remarks and
comments on the paper. The paper was partly supported by the Polish KBN
grant 2P03D00814 to  A.\ Udalski. Partial support for the OGLE project
was provided with the NSF  grant AST-9530478 to B.~Paczy\'nski.}

\end{document}